\documentclass[12pt]{iopart}
\usepackage{graphicx}
\begin{document}

\article[Dynamics of triggered inclusions]{Paper}{On the dynamics of the interaction between triggered active inclusions}

\author{Jean-Baptiste Fournier and K\'evin Sin Ronia}
\address{Universit\'e Paris Diderot, Sorbonne Paris Cit\'e, Laboratoire Mati\`ere et Syst\`emes Complexes (MSC), UMR 7057 CNRS
F-75205 Paris, France, EU}
\ead{jean-baptiste.fournier@univ-paris-diderot.fr}

\begin{abstract}
In a one-dimensional elastic medium with finite correlation length and purely relaxational dynamics, we calculate the time dependence of the elastic force $\mathcal{F}(t)$ exchanged between two active inclusions that trigger an elastic deformation at $t=0$. We consider (i) linear inclusions coupling to the field with a finite force, and (ii) non-linear inclusions imposing a finite deformation. In the non-linear case, the force exhibits a transient maximum much larger than the equilibrium force, diverging as $\sim\!L^{-2}$ at separations $L$ shorter than the field's correlation length. Both the mean-field and the Casimir component of the interaction are calculated. We also discuss the typical appearance time and equilibration time of the force, comparing the linear and the non-linear cases. The existence of a high transient force in the non-linear case should be a generic feature of elastically-mediated interactions.
\end{abstract}

\noindent{\it Keywords\/}: Statistical mechanics of complex materials: stochastic processes (theory), Non-equilibrium processes: exacts results,  Interface biology-physics: active membranes

\submitto{J.\ Stat.\ Mech.}

\maketitle


\section{Introduction}

Inclusions placed in an elastic medium usually deform it, and experience mediated interactions as the total elastic free energy depends on their separation. This occurs in crystals and in soft matter systems, such as capillary interfaces, bilayer membranes, liquid crystals, and binary mixtures~\cite{Lubensky_book,Safran_book}. In systems exhibiting near-critical fluctuations, a Casimir-like effect adds up, which is due to the coupling between the boundary conditions and the fluctuations of the medium~\cite{Fisher78,Gambassi09}. While these effects are rather well understood, the \textit{dynamics} of elastically--mediated interactions remains a challenging subject. Mobile inclusions couple to the dynamics of the elastic medium, which produces non trivial effects~\cite{Rafii05,Demery10,Chen10,Dean11}, as in the dynamics of fluctuation-induced forces~\cite{Bartolo03,Rodriguez-Lopez11,Dean12}

Here, we investigate the dynamics of the force exchanged between two fixed but active inclusions, that are simultaneously triggered. For $t<0$, the inclusions impose no deformation to the medium, either they are outside the medium, or they are in a state in which they do not couple with its elastic field. At $t=0$, they are triggered, i.e.~they actively change state or insert into the medium, and they start to impose an elastic deformation. We consider two limiting cases: (i) soft, linear inclusions, and (ii) hard, non-linear inclusions. In the former case, the inclusions couple linearly to the elastic field with a \textit{constant force}. It is then the competition with the medium that sets the amplitude of the deformation. In the latter case, the inclusions couple non-linearly with the elastic field in such a way that they impose a \textit{constant deformation}. In both cases, we assume that the time scale associated with the switching of the inclusions is much shorter than the response time of the system; in other words the switching of the  inclusions is assumed to be instantaneous. These are realistic situations: membrane proteins, for instance, may switch upon binding of ATP from a non-curving cylindrical shape to a curving conical one~\cite{Goulian93}. Either the protein opens-up with a finite force, or it abruptly changes its shape and imposes a local membrane curvature. For membranes proteins, the latter case is more realistic, as proteins are much stiffer that the surrounding membrane. Assuming an instantaneous shape change is also a very good approximation, since the dynamics of the membrane is much slower than that of the protein at distances comparable with inter-protein distances~\cite{Seifert93}. 

\section{Soft, linear inclusions}

For the sake of simplicity, we consider a one dimensional elastic medium described by a scalar Gaussian field $\phi(x,t)$, and a purely dissipative dynamics of the model A type.
The Hamiltonian $\mathcal{H}$ of the system is equal to the elastic medium's Hamiltonian:
\begin{equation}
\mathcal{H}_\mathrm{el}=\int\rmd x\left\{
\frac{1}{2}r\,\phi(x,t)^2+\frac{1}{2}c\left[\nabla\phi(x,t)\right]^2\right\},
\end{equation}
plus the Hamiltonian of the inclusions. For soft, linear inclusions, triggered at $t=0$, we take for the latter:
\begin{equation}
\mathcal{H}_\mathrm{inc}=B\,\theta(t)\int\rmd x\left[
\delta(x)\phi(x,t)+\delta(x-L)\phi(x,t)
\right],
\end{equation}
the inclusions being placed at $x=0$ and $x=L$.
Here, $\theta(t)$ is the Heaviside step function, $\delta(t)$ is the Dirac distribution, $B$ is the strength of the inclusions, and $\nabla=\partial_x$. Assuming a purely relaxational dynamics, the time evolution of the field is given by~\cite{Zwanzig_book}
\begin{eqnarray}
\fl\qquad
\Gamma^{-1}\,\partial_t\phi(x,t)
&=-\frac{\delta\mathcal{H}}{\delta\phi(x,t)}+\eta(x,t)
\nonumber\\
&=\left(c\nabla^2-r\right)\phi(x,t)-B\theta(t)\delta(x)-B\theta(t)\delta(x-L)
+\eta(x,t)\,,
\end{eqnarray}
where $\eta(x,t)$ is a thermal Gaussian noise satisfying $\langle\eta(x,t)\,\eta(x',t')\rangle=2\Gamma^{-1}k_\mathrm{B}T\delta(x-x')\delta(t-t')$, with $T$ the temperature and $k_\mathrm{B}$ Boltzmann's constant.

Let us rescale all lengths by the field's correlation length $\xi=(c/r)^{1/2}$, all times by $\tau=(\Gamma r)^{-1}$, and all energies by $\epsilon=r\xi=(rc)^{1/2}$, so that everything becomes dimensionless. The dynamical equation and the noise correlation function become:
\begin{eqnarray}
\label{eqdyn}
\fl\qquad\quad
\partial_t\phi(x,t)=\left(\nabla^2-1\right)\phi(x,t)-B\theta(t)\delta(x)-B\theta(t)\delta(x-L)+\eta(x,t)\,,
\\
\fl\qquad\quad
\langle\eta(x,t)\,\eta(x',t')\rangle=2T\delta(x-x')\delta(t-t')\,.
\end{eqnarray}
Note that $B$ has been rescaled by $\epsilon$ and $T$ by $\epsilon/k_\mathrm{B}$.

\subsection{Force exchanged between the inclusions}

The interaction between the inclusions can be computed from the stress-tensor associated to the elastic medium~\cite{Landau_book,Bartolo03,Fournier08,Bitbol10} (see also the discussions in Refs.~\cite{Bitbol11,Dean12}). With the normalized Hamiltonian density $h_\mathrm{el}=\frac{1}{2}\phi^2+\frac{1}{2}(\nabla\phi)^2$, the stress tensor is given by
\begin{equation}
\sigma(x,t)=h_\mathrm{el}-(\nabla\phi)\,\partial h_\mathrm{el}/\partial(\nabla\phi)
=\frac{1}{2}\phi^2(x,t)-\frac{1}{2}[\nabla\phi(x,t)]^2\,.
\end{equation}
Hence, the ensemble average of force acting on the inclusion at $x=0$ is given by
\begin{equation}
\label{force}
\mathcal{F}(t)=\langle\sigma(0^+,t)-\sigma(0^-,t)\rangle=
\langle\frac{1}{2}[\nabla\phi(0^-,t)]^2\rangle
-\langle\frac{1}{2}[\nabla\phi(0^+,t)]^2\rangle\,.
\end{equation}
Note that the $\phi^2$ terms have canceled out because of the continuity of $\phi$ in $x=0$.

Let us decompose the field as its ensemble average plus its stochastic component:
\begin{equation}
\phi(x,t)=\langle\phi(x,t)\rangle+\tilde\phi(x,t)=\Phi(x,t)+\tilde\phi(x,t)\,.
\end{equation}
Since $\langle\tilde\phi(x,t)\rangle=0$, the force $\mathcal{F}(t)$ exchanged between the inclusions can be written as $\mathcal{F}(t)=F(t)+F_\mathrm{C}(t)$, with $F(t)$ its the mean-field component and $F_\mathrm{C}$ it's Casimir, fluctuation-induced, component:
\begin{eqnarray}
\label{Fcm}
F(t)=\frac{1}{2}\left[\nabla\Phi(0^-,t)\right]^2
-\frac{1}{2}\left[\nabla\Phi(0^+,t)\right]^2,
\\
\label{FC}
F_\mathrm{C}(t)=\langle\frac{1}{2}[\nabla\tilde\phi(0^-,t)]^2\rangle
-\langle\frac{1}{2}[\nabla\tilde\phi(0^+,t)]^2\rangle.
\end{eqnarray}

Let us first deal with the Casimir component $F_\mathrm{C}^\mathrm{\,soft}(t)$. Because the theory is Gaussian, and because $\mathcal{H}_\mathrm{inc}$ contains only terms that are linear in $\phi$, the correlation function $C(x,x';t,t')=\langle\tilde\phi(x,t)\,\tilde\phi(x',t')\rangle$ of the stochastic part $\tilde\phi(x,t)$ of the field is the same as in the case $B=0$. Using the Martin--Siggia--Rose formalism~\cite{Hohenberg77}, one can show that it obeys the equation
\begin{equation}
\label{MSR}
\left(-\partial_t-\nabla^2+1\right)\left(\partial_t-\nabla^2+1\right)
C(x,x';t,t')=2T\delta(x-x')\delta(t-t'),
\end{equation}
at all points of space and time. As the physics is translationally invariant for $\tilde\phi$ (in the soft, linear case), the quantity $\langle[\nabla\tilde\phi(x,t)]^2\rangle$ is space independent, therefore continuous, and
\begin{equation}
F_\mathrm{C}^\mathrm{\,soft}(t)=0.
\end{equation}
There is no fluctuation-induced component of the force in the case (i) of soft, linear inclusions.

Before studying the mean-field component $F(t)$, let us study the time evolution of $\Phi(x,t)=\langle\phi(x,t)\rangle$.
Taking the ensemble average of equation~(\ref{eqdyn}) then its Laplace transform, with $\hat\Phi(x,s)=\int_0^\infty\rmd t\, \Phi(x,t)\exp(-st)$, we obtain
\begin{equation}
\label{eqdyn2}
\nabla^2\hat\Phi(x,s)-(1+s)\hat\Phi(x,s)=\frac{B}{s}\left[\delta(x)+\delta(x-L)\right].
\end{equation}
The solution is
\begin{equation}
\label{sol}
\fl\qquad
\hat\Phi(x,s) = 
\cases{
  C_1\exp\!\left(x\sqrt{1+s}\right) & for $x\le 0$\,, \cr
  C_2\left\{\exp\!\left(-x\sqrt{1+s}\right)+\exp\!\left[(x-L)\sqrt{1+s}\right]\right\} & for $0\le x\le L$\,, \cr
  C_1\exp\!\left[-(x-L)\sqrt{1+s}\right] & for $x\ge L$\,,
}
\end{equation}
with $C_2=B/(2s\sqrt{1+s})$ and $C_1/C_2=1+\exp(-L\sqrt{1+s})$. The asymptotic deformation set by the inclusions is $\Phi_0\equiv\lim_{t\to\infty}\Phi(0,t)=\lim_{s\to0}[s\hat\Phi(0,s)]=\frac{1}{2}B[1+\exp(-L)]$. We choose $B=2/[1+\exp(-L)]$ in order to normalize the field in such a way that $\Phi_0=1$. We thus obtain $1/C_2=[1+\exp(-L)]s\sqrt{1+s}$, which completely defines $\hat\Phi(x,s)$.

\begin{figure}
\centerline{\includegraphics[width=\columnwidth]{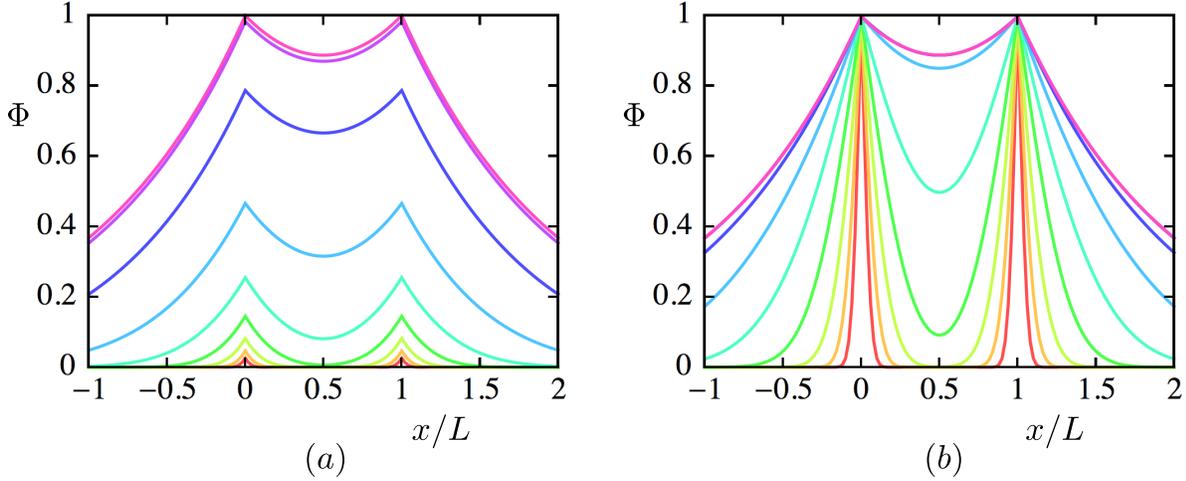}}
\caption{Time evolution of the elastic deformation $\Phi(x)=\langle\phi(x,t)\rangle$ created by two inclusions placed in $x=0$ and $x=L$, and triggered at $t=0$. (a) soft, linear inclusions imposing a constant force. (b) hard, non-linear inclusions imposing a constant deformation. From bottom to top the time increases logarithmically, with values $t=0.001$, $t=0.00316$, $t=0.01$, $t=0.0316$, $t=0.1$, $t=0.316$, $t=1$, $t=3.16$ and $t=10$.}
\label{temporel}
\end{figure}

Figure~\ref{temporel}$a$ shows the profiles of $\Phi(x,t)$ (in the soft, linear case), obtained by numerically inverting the Laplace transform using Durbin's method~\cite{Durbin74}. At short times, the inclusions are not aware of each other, and the deformation is approximatively symmetrical around each inclusion, displaying a characteristic conical shape with fixed angle. Indeed, the term $-(B/s)\delta(x)$ in the dynamical equation implies $\nabla\hat\Phi(0^+,s)\simeq-\frac{1}{2}B/s$ (by symmetry), which yields a constant slope $\nabla\Phi(0^+,t)\simeq-\frac{1}{2}B$. Taking the limit $L\to\infty$, as the inclusions ignore each other at short times, the deformation obeys $\hat\Phi(0,s)=C_1\simeq C_2\sim s^{-3/2}$ for $s\to\infty$, which implies $\Phi(0,t)\sim t^{1/2}$. With the constant slope, this implies that the deformation spreads as $\Delta x\sim t^{1/2}$, as expected for purely relaxational dynamics. We therefore expect that the inclusions will start interacting after a time $\Delta t\sim L^2$.

Let us now calculate the force $F(t)$ by using equation~(\ref{Fcm}). The quantities $\nabla\Phi(0^\pm,t)$ contributing to $F$ can be calculated analytically by computing the inverse Laplace transform of $\nabla\hat\Phi(0^\pm,s)=[\exp(-L\sqrt{1+s})\mp1]/[s(1+\exp(-L)]$. We thus obtain for the mean-field force:
\begin{equation}
\label{forcemou}
\bar F^\mathrm{\,soft}(t)=\frac{F^\mathrm{\,soft}(t)}{F_\mathrm{eq}}=
\frac{1}{2}\left[
\mathrm{erfc}\!\left(\frac{L-2t}{2\sqrt{t}}\right)
+\rme^{2L}\,\mathrm{erfc}\!\left(\frac{L+2t}{2\sqrt{t}}\right)
\right],
\end{equation}
where
\begin{equation}
F_\mathrm{eq}=\lim_{t\to\infty}F^\mathrm{\,soft}(t)=\left[1+\cosh(L)\right]^{-1}
\end{equation}
is the equilibrium force. Both quantities being positive, the interaction between the inclusions is attractive.
Note that $F_\mathrm{eq}$ can easily be obtained by solving the static problem. It is obviously the common asymptotic limit of $F(t)$ both in the case (i) of soft, linear inclusions and in the case (ii) of hard, non-linear inclusions, as the asymptotic profiles are identical.

\begin{figure}
\centerline{\includegraphics[width=\columnwidth]{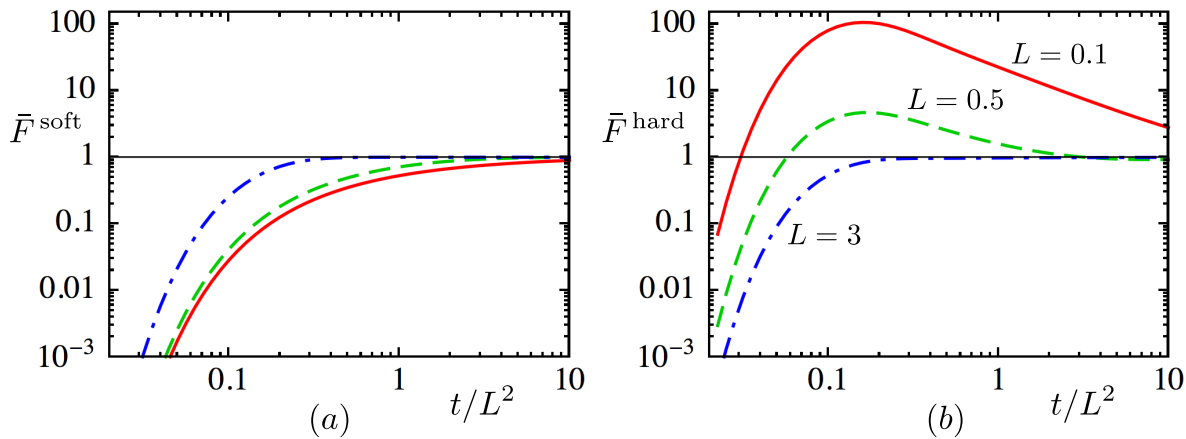}}
\caption{Normalized attractive interaction~$\bar F$ between the inclusions as a function of the elapsed time~$t$ scaled by their squared separation~$L^2$. (a) soft, linear inclusions, (b) hard, non-linear inclusions. Plain (red) curves: $L=0.1$. Dashed (green) curves: $L=0.5$. Dash-dotted (blue) curves: $L=3$.}
\label{forces}
\end{figure}

The spreading of the deformation at short times as $\Delta x\sim t^{1/2}$ suggests to plot $\bar F(t)$ as a function of $t/L^2$. Figure~\ref{forces}$a$ shows that this scaling is reasonably good, although there is no true scale invariance in this problem (due to the existence of a characteristic length and a characteristic time, both equal to unity here).

\section{Hard, non-linear inclusions}

We now set $\mathcal{H}_\mathrm{inc}=0$ (i.e.~$B=0$) and we impose the following ``hard" boundary conditions: $\phi(0,t)=\phi(L,t)=1$, $\forall t\ge0$. Note that for $t<0$ the field fluctuates freely with no boundary conditions. Because $\mathcal{H}_\mathrm{el}$ is unchanged, and because we still use the purely relaxational model A dynamics, the time evolution of $\phi(x,t)$ is still given by~(\ref{eqdyn}), with $B=0$, and its ensemble average $\hat\Phi(x,s)$ by~(\ref{eqdyn2}), also with $B=0$. With the hard boundary conditions in Laplace form, this yields
\begin{eqnarray}
\nabla^2\hat\Phi(x,s)-(1+s)\hat\Phi(x,s)=0\,,\\
\hat\Phi(0,s)=\hat\Phi(L,s)=1/s\,,\quad\forall s\,.
\end{eqnarray}
The solution is still given by (\ref{sol}), but the boundary conditions now yield $C_1=1/s$ and $1/C_2=[1+\exp(-L\sqrt{1+s})]s$. Figure~\ref{temporel}$b$ shows the profiles of $\Phi(x,t)$, obtained by numerically inverting the Laplace transform using Durbin's method. Again, at short times, the inclusions are not aware of each other, and the deformation around each inclusion is symmetrical. The slope of the deformation profile is given by $\nabla\hat\Phi(0^-,s)=\sqrt{1+s}\,C_1\sim s^{-1/2}$ as $s\to\infty$. Hence, at short times $\nabla\hat\Phi(0^-,t)\sim t^{-1/2}$, and since $\Phi(0,t)=1$, the deformation spreads again diffusively as $\Delta x\sim t^{1/2}$.

The force exchanged between the inclusions is still given by $\mathcal{F}(t)=F(t)+F_\mathrm{C}(t)$, with $F(t)$ given by (\ref{Fcm}) and $F_\mathrm{C}(t)$ given by~(\ref{FC}), as the elastic Hamiltonian is unchanged. Let us first discuss the mean-field force $F^\mathrm{\,hard}(t)$. Contrary to the case (i) of soft, linear inclusions, there is no analytical solution, hence we used Durbin's numerical method to compute the inverse Laplace transforms of $\nabla\hat\Phi(0^\pm,s)$. Figure~\ref{forces}$b$ shows the time evolution of $\bar F^\mathrm{\,hard}(t)=F^\mathrm{\,hard}(t)/F_\mathrm{eq}$ for the same set of separations $L$ as in the soft case.

\begin{figure}
\centerline{\includegraphics[width=\columnwidth]{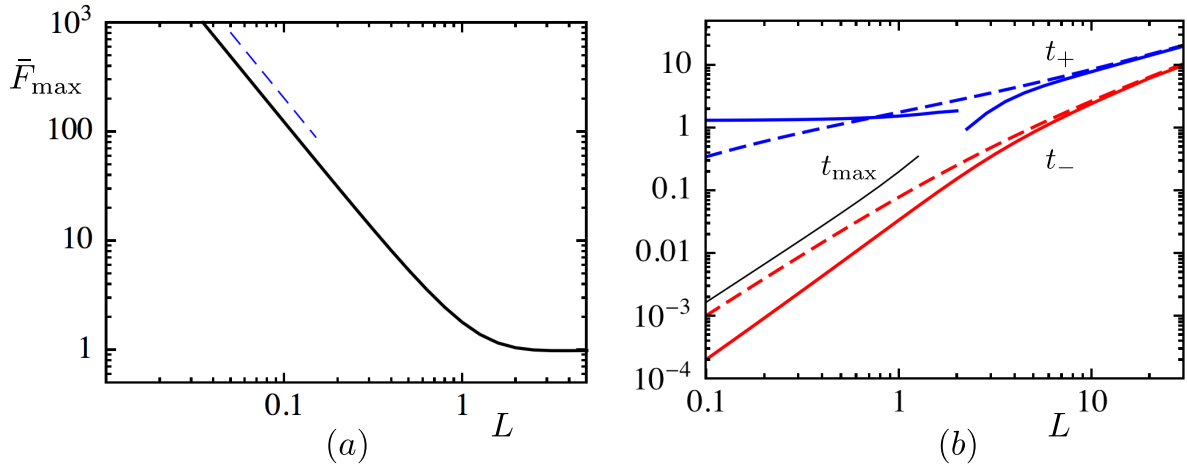}}
\caption{(a) Maximum of $\bar F(t)$, in the case of hard, non-linear inclusions, as a function of the inclusions separation~$L$. The dotted (blue) line has a slope of $2$. (b) Appearance time $t_-$ (lower red curves), equilibration time $t_+$ (upper blue curves), and maximum time $t_\mathrm{max}$ (middle black curve) of the force $F(t)$, for a convergence criterion of $p=3\%$, as a function of the inclusions separation $L$. Dashed curves: soft, linear inclusions; plain curves: hard, non-linear inclusions.}
\label{last}
\end{figure}

The striking feature is the existence of a bump, which becomes very large when the inclusions are at a distance shorter than the field's correlation length (i.e.\ $L<1$). As shown in figure~\ref{last}$a$, the maximum of the force actually diverges as
\begin{equation}
\bar F_\mathrm{max}^\mathrm{\,hard}\sim L^{-2}\,,\quad\mathrm{for~}L\to0.
\end{equation}
Indeed, when the deformations produced by the inclusions start to merge, the gradient of $\Phi$ is of order $1/L$, and asymmetric (see figure~\ref{temporel}$b$), which by (\ref{force}) implies the scaling of the force. This reasoning holds only if the deformations have not relaxed spatially when they merge, which implies the disappearance of the maximum when $L$ reaches unity (figure~\ref{last}$a$).

The Casimir component $F_\mathrm{C}^\mathrm{\,hard}(t)$ is given by~(\ref{FC}), and it can be derived from the correlation function. Using again the Martin--Siggia--Rose formalism, we find that the correlation function $C(x,x';t,t')=\langle\tilde\phi(x,t)\,\tilde\phi(x',t')\rangle$ is still given by~(\ref{MSR}), as in the soft case (i), but now with boundary conditions such that $C(x,x';t,t')$ must vanish if either $x$ or $x'$ is equal to $0$ or $L$ at positive times. We thus have Dirichlet conditions in $x=0$ and $x=L$ for the fluctuating field $\tilde\phi$ at positive times, while there are no boundary conditions at negative times. There is no need, actually, to calculate $C(x,x';t,t')$. Since the model is Gaussian, the Fourier modes $\tilde\phi(q,t)$ are independent. They are thermally equilibrated at $t<0$, then, at $t=0$ all the modes that do not satisfy the Dirichlet boundary conditions are suddenly removed while the other remain thermally equilibrated. It follows that the Casimir interaction jumps abruptly from 0 to its equilibrium value:
\begin{equation}
F_\mathrm{C}^\mathrm{\,hard}(t)=\frac{T}{e^{2L}-1}\theta(t)\,.
\end{equation}
Note that this situation differs from that discussed in~\cite{Dean10}, where the dynamics of the Casimir force is calculated for a system prepared in the frozen state $\tilde\phi=0$ at time $t=0$. To calculate the equilibrium value of the Casimir force, we have used the method of Li and Kardar~\cite{Li92}: the static correlation function of the field free of boundary conditions is $G(x)=\frac{1}{2}e^{-|x|}$, which yields the Casimir energy $\frac{1}{2}T\ln[1-G^2(L)/G^2(0)]$ and the Casimir force $T/(e^{2L}-1)$ by differentiation with respect to~$L$. Note that $F_\mathrm{C}^\mathrm{\,hard}(t)$ always overcomes the mean-field contribution $F^\mathrm{\,hard}(t)$ at small enough separations, but their detailed comparison depend on~$T$.

\section{Dynamical regimes of the force}

To complete the dynamical study of the mean-field force $\bar F(t)$, let us define the \textit{appearance} time $t_-$, and the \textit{equilibration} time $t_+$, by the conditions
\begin{eqnarray}
\forall t\ge t_-\,,\quad\bar F(t)\ge p\,;\\
\forall t\ge t_+\,,\quad\left|1-\bar F(t)\right|\le p\,,
\end{eqnarray}
with $p\ll1$. The behaviors of $t_-(L)$ and $t_+(L)$ are shown in figure~\ref{last}$b$ for $p=3\%$ (i.e.~a few percent).

Let us first discuss the case (i) of soft, linear inclusions (dashed curves in figure~\ref{last}$b$). For $L\ll1$, we expect scaling laws, as the characteristic lengths and times are irrelevant. Indeed, in this regime the force is well approximated by $\bar F(t)\simeq \mathrm{erfc}(\frac{1}{2}L/\sqrt{t})$, which can be deduced from (\ref{forcemou}) by setting $\exp(2L)\simeq 1$ and $t\ll L$. It follows that
\begin{eqnarray}
t_\pm\simeq L^2/D_\pm\,,\qquad \mathrm{for~}L\ll1\,,
\end{eqnarray}
with
\begin{equation}
D_-=\left[2\,\mathrm{erfc}^{-1}(p)\right]^2,
\quad
D_+=\left[2\,\mathrm{erfc}^{-1}(1-p)\right]^2\simeq\pi p^2.
\end{equation}
For $p=3\%$ one obtains $D_-\simeq9.4$ and $D_+\simeq2.3\times10^{-3}$. Note that in figure~\ref{last}$b$ the regime $t_-\sim L^2$ is apparent, while the regime $t_+\sim L^2$ occurs at values of $L$ smaller than those displayed. In the opposite regime $L\gg1$, where there are no scaling laws, the force at large times is well approximated by the first term of (\ref{forcemou}), i.e.~$\bar F(t)\simeq\frac{1}{2}\mathrm{erfc}[\frac{1}{2}(L-2t)/\sqrt{t}]$. It follows that
\begin{equation}
\label{tapeqLgrand}
t_\pm\simeq\frac{1}{2}\left(L+\alpha\pm\sqrt{2\alpha L+\alpha^2}\right),
\qquad \mathrm{for~}L\gg1\,,
\end{equation}
where
\begin{equation}
\alpha=\left[\mathrm{erfc}^{-1}(2p)\right]^2\,,
\end{equation}
For $p=3\%$ one obtains $\alpha\simeq1.8$. The establishment of the force therefore lasts $t_+-t_-\simeq\sqrt{\alpha(2L+\alpha)}$ in this regime.

Let us now turn to the case (ii) of hard, non-linear inclusions. First of all, for $L\gg1$, we notice on figure~\ref{last}$b$ that the curves $t_\pm(L)$ are superimposed on the dashed ones of case (i). They are therefore given by~(\ref{tapeqLgrand}). Indeed, whatever the inclusion type, the deformation profiles are already well established in this regime when their tails start to overlap. In the range of values $L\ll1$ that we have explored, $t_-(L)$ is well fitted numerically by a power law $L^\lambda$; we found $\lambda\simeq2.15\pm0.05$ (instead of $\lambda=2$), which seems to be independent of $p$ in the range $[0.1\%,10\%]$. A more detailed numerical study would be required, however, to discuss seriously this scaling law. Conversely, the time $t_\mathrm{max}$ at which the force maximum occurs is found numerically to follow an analytic power-law:
\begin{equation}
t_\mathrm{max}\simeq L^2/D_\mathrm{max}\,,
\end{equation}
with $D_\mathrm{max}\simeq6.0$.
As for the equilibration time, we find numerically $t_+(L)$ of order unity for $L<1$, with $t_+(L)\to t_0\simeq1.3$ as $L\to0$. In other words, the equilibration time is set by the field's own timescale, even when the inclusions are very close to each other. Another characteristic feature of hard, non-linear inclusions is the existence of a discontinuity in $t_+(L)$, occuring at $L\simeq2.25$ for $p=3\%$ (see figure~\ref{last}$b$). This discontinuity comes from the maximum of the force $\bar F(t)$: if the latter is larger than $1+p$ then $t_+(L)$ lies after the maximum, otherwise it lies before the maximum. Overall, for $L<1$, the force in the hard inclusions case appears sooner and equilibrates later than in the soft inclusions case.

\section{Conclusion}

We showed that the mechanism by which active, switchable inclusions set their deformations in an elastic medium has important consequences on the dynamics of the force they exchange. In the example studied here, two inclusions instantaneously triggering their deformations may experience a transient mean-field force that can be hundreds of times larger than the equilibrium one. Inclusions setting their deformations through an instantaneous force do not exhibit such a behavior. The timescales associated with the dynamics of the force are also largely affected by their switching mechanism. The existence of a high transient force for inclusions setting their deformation instantaneously (i.e.~much quicker than the deformation of the medium) should be a generic feature of elastically mediated interactions. It might have important consequences. For instance, two active inclusions tied to the medium by a link withstanding the equilibrium force could either break free, or remain attached, depending on the nature of their switching mechanism.
\ack

We thank F. van Wijland for helpful interactions and critical reading of our manuscript.


\end{document}